\documentclass[prl,twocolumn, final, showpacs, floatfix]{revtex4}
\usepackage{amssymb,amsmath}
\usepackage{graphicx}

\usepackage[scanall]{psfrag}

\usepackage{color}

\newcommand{\psLabelSize}{1.3}
\newcommand{\psLabelSz}{0.9}
\newcommand{\psSymbolSize}{0.8}

\newcommand{\abs}[1]{\left| #1 \right|}

\begin{document}

\title{Chiral Surfaces Self-Assembling in One-Component Systems with Isotropic Interactions}

\author{E. Edlund}
\author{O. Lindgren}
\author{M. \surname{Nilsson Jacobi}}%
 \email{mjacobi@chalmers.se} 
\affiliation{Complex Systems Group, Department of  Energy and Environment, Chalmers University of Technology, SE-41296 G\"oteborg, Sweden}
\date{\today}%

\begin{abstract}
We show that chiral symmetry can be broken spontaneously in one-component systems with isotropic interactions, i.e. many-particle systems having maximal a priori symmetry. This is achieved by designing isotropic potentials that lead to self-assembly of chiral surfaces. We demonstrate the principle on a simple chiral lattice and on a more complex lattice with chiral super-cells. In addition we show that the complex lattice has interesting melting behavior with multiple morphologically distinct phases that we argue can be qualitatively predicted from the design of the interaction.
\end{abstract}


\maketitle


Breaking of chiral symmetry plays a central role both in fundamental physics, e.g., parity violation in the weak interaction, and in biology where many types of biomolecules exist only as enantiomers. The commonly accepted explanation for homochirality in chemical and biological systems is spontaneous symmetry breaking. 
This phenomenon involves two steps: first the formation of chiral molecules, and then a chiral specific catalysis that amplifies a stochastic imbalance to a macroscopic scale. In heterogeneous systems it is not hard to imagine that both steps can be achieved, and there are indeed many chemical systems that spontaneously deviate from racemic mixture~\cite{okamoto94,eliel94}. 
There are even several known molecules that show chiral auto-catalytic activity~\cite{soai95,Soai02}. It is an interesting question whether spontaneous breaking of chiral symmetry can happen also in simpler systems, by which we here mean systems with higher degree of symmetry. 
The perhaps simplest such class of systems would consist of a single  particle type with an isotropic pairwise interaction. For several reasons it appears to be much harder to form homochiral states in such systems. 
First, cluster formation, which mimics the creation of molecules, typically results in achiral structures with dihedral group symmetries~\cite{edlund_novel_2011}. Second, (auto-) catalysis is not easily achieved in homogeneous isotropic systems. 
However, chiral symmetry can also be broken in another context, namely during crystallization. 
Again there are many examples of heterogeneous systems where this is known to happen~\cite{kondepudi90,Flack03}, but no chiral crystals arising from isotropic interactions have been reported, neither in explorative~\cite{engel_self-assembly_2007} nor design~\cite{torquato_inverse_2009} studies. 
In this Letter we show that one-component systems with carefully designed isotropic interactions can self-assemble into chiral lattices. Our results demonstrate that spontaneous breaking of chiral symmetry can occur in many-particle systems with maximal degree of symmetry. 

Chiral surfaces are of practical importance because of their potential for use in chiral catalysis.  The most important applications are in the pharmaceutical industry where we learned the hard way that a therapeutically well behaved enantiomer may be toxic in the other enantiomeric form. 
The classic example is the tragedy with Thalidomide, whose racemic form turned out to cause birth defects after being extensively used as a sedative for pregnant women~\cite{blaschke_chromatographic_1979}. 
Today the dominating method for synthesizing large quantities of chiral products involves various types of chiral catalysts in solution (homogeneous catalysis)~\cite{yoon_privileged_2003, jacobsen_comprehensive_1999}. However, the development of chiral surface  catalysts (heterogeneous catalysis) receive much interest due to their  separability and reusability. 
Several ways of producing chiral surfaces exist, for example through surface reconstruction induced by adsorbed chiral molecules~\cite{schunack_anchoring_2001}, cleaving achiral crystals along planes of low symmetry such as the (643) surface of an fcc structure~\cite{sholl_naturally_2001}, or self-assembly of molecules into chiral structures on achiral surfaces~\cite{ortega_lorenzo_extended_2000,spillmann_hierarchical_2003}. 
One of the outstanding challenges for all heterogeneous chiral catalysis is how to make chiral surfaces with sufficiently large (macroscopic) active areas~\cite{gellman_chiral_2010}. Simple models that exhibit formation of chiral domains, such as the models presented here, could prove useful in supporting progress in this area. 

\begin{figure}[tb]
\centering
\psfrag{a}[c][c][\psLabelSize][0]{\color[rgb]{1,1,1}\bf{a}}
\psfrag{b}[c][c][\psLabelSize][0]{\color[rgb]{1,1,1}\bf{b}}
\psfrag{c}[c][c][\psLabelSize][0]{\color[rgb]{1,1,1}\bf{c}}
\psfrag{d}[c][c][\psLabelSize][0]{\color[rgb]{1,1,1}\bf{d}}
\psfrag{U}[r][c][\psSymbolSize][0]{$\widehat{V}(k)$}
\psfrag{k}[c][c][\psSymbolSize][0]{$k$}
\psfrag{V}[r][c][\psSymbolSize][0]{$V(r)$}
\psfrag{r}[c][c][\psSymbolSize][0]{$r$}
\psfrag{A}[c][c][\psSymbolSize][0]{$l_1$}
\psfrag{B}[c][c][\psSymbolSize][0]{$l_2$}
\psfrag{C}[c][c][\psSymbolSize][0]{$l_3$}

\includegraphics[width=0.45 \textwidth]{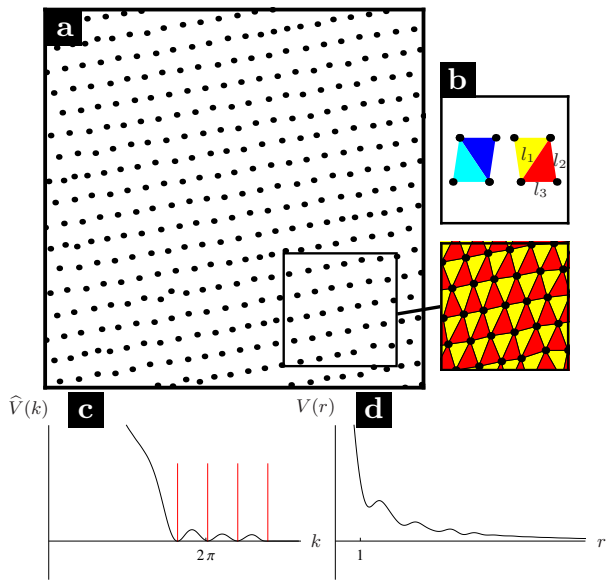}
\caption{ \label{paralellogram} 
(a) The simplest form of chiral lattice, composed of scalene (acute) triangles with well separated side lengths $l_i$. The emerging chirality can be either left or right oriented (b). (c) By selecting for the reciprocal lattice (the red peaks) of the target structure, a potential (d) which causes self-assembly into one of the two possible chiralities is obtained. 
}
\end{figure}
\begin{figure}[tb]
\centering
\psfrag{a}[c][c][\psLabelSize][0]{\color[rgb]{1,1,1}\bf{a}}
\psfrag{b}[c][c][\psLabelSize][0]{\color[rgb]{1,1,1}\bf{b}}
\psfrag{c}[c][c][\psLabelSize][0]{\color[rgb]{1,1,1}\bf{c}}
\psfrag{d}[c][c][\psLabelSize][0]{\color[rgb]{1,1,1}\bf{d}}
\psfrag{U}[r][c][\psSymbolSize][0]{$\widehat{V}(k)$}
\psfrag{k}[c][c][\psSymbolSize][0]{$k$}
\psfrag{V}[r][c][\psSymbolSize][0]{$V(r)$}
\psfrag{r}[c][c][\psSymbolSize][0]{$r$}
\psfrag{R}[c][t][\psSymbolSize][0]{\color[rgb]{1,0,0}$\hat{\rho}_{max}$}
\psfrag{e}[r][c][\psSymbolSize][0]{$-\epsilon$}
\psfrag{P}[c][b][0.7][0]{$2\pi$}

\includegraphics[width=0.45 \textwidth]{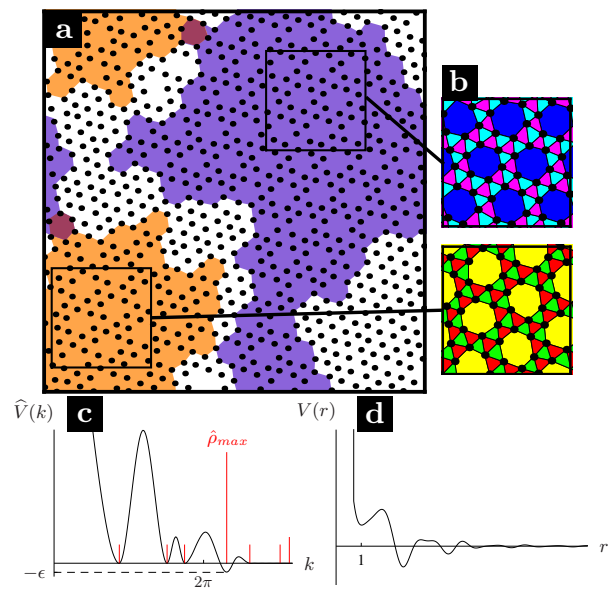}
\caption{ \label{snubHex} 
Snub hexagonal tiling, a chiral Archimedean tiling. Note that the chirality does not originate from the positions of the hexagons (they are organized in rombs and not parallelograms as they must when the tiles are regular polygons) but from the orientations of the triangles in between.
(a-d) as in Fig.~\ref{paralellogram}. In (a) two regions of opposite chirality are highlighted. (c) illustrates the perturbation of magnitude $\epsilon$ at the maximum of $\hat{\rho}$.
}
\end{figure}

In this work we focus on self-assembly at low temperature where structure formation is driven by minimization of the potential energy. For a given configuration in a system with isotropic interactions, the energy is defined by the density distribution $\rho({\bf r})$ and is naturally described as a quadratic form that sums all the contributions from the pairwise potential. For our purposes it is suitable to express the energy in reciprocal space where the quadratic form is diagonalized due to the translational invariance of the isotropic interactions~\cite{edlund_universality_2010}: 
\begin{eqnarray} \label{fourierEnergy}
	E & = &  \int \! \mathrm{d} {\bf r}_1 \mathrm{d} {\bf r}_2 \, \rho ({\bf r}_1 ) V ( | {\bf r} _1 -  {\bf r} _2 | ) \rho ({\bf r}_2 ) \nonumber \\
	& = & 2\pi \int \! \mathrm{d}k \,\abs{\hat{\rho}(k)}^2 \widehat{V}(k) , 
 \end{eqnarray} 
 where $|\hat{\rho} (k)|^2  \mathrm{d}k= \int _{ |s{\bf k}|= k } \mathrm{d} {\bf k} \, |\hat{\rho} ( {\bf k} )|^2$, $\hat{\rho} ( {\bf k} )$ is the standard Fourier transform of $\rho ( {\bf r} )$, and $\widehat{V}(k)$ is the radial Fourier (Hankel) transform of $V(r)$~\cite{folland_fourier_2009}, $\widehat{V}(k)=\int \! r  \mathrm{d} r \, V(r) J_0 (r k)$. 
The small $|k|$ region describes the defining features of the crystal structure and by designing the energy spectrum of this region particles can be made to self-assemble into crystalline structures at low temperatures. We used this observation in a recent study to show how to design the interactions of a system so that it self-assembles into target lattices~\cite{edlund_designing_2011}. 
Briefly, the method works as follows. The symmetry of the lattice manifests in reciprocal space as a restriction of the support of $\hat{\rho}({\bf k})$ to a set of finite points $\{{\bf G}_i\}$. By choosing $\widehat{V}(k)$ smooth and positive with zeros coinciding with the reciprocal lattice 
${\bf G}_i$ of the target structure $\widehat{V}(|{\bf G}_i|)=0$ [see Figs.~\ref{paralellogram}(c) and \ref{snubHex}(c)], we can guarantee the target configuration to be a ground state. This construction utilizes the fact that there are a finite number of Bravais lattices, all with different structure factors. The structure factors, $\hat{\rho}({\bf G}_i)$, will not affect the energy and hence all crystals with the same periodicity will be ground states.



Chirality on a crystal surface is by definition equivalent to the absence of axes of symmetry and it can manifest itself in fundamentally different ways. Every ideal crystal structure consists of a mathematical lattice and a basis.~\footnote{We have up to now followed the common but inexact usage of the word lattice to mean any ideal crystal structure, but in the following three paragraphs we need to be more exact.} If either the basis or the lattice is chiral, the crystal will be as well. But even if both lattice and basis are achiral, the crystal can be chiral if the axes of symmetry of the lattice does not coincide with those of the basis. In this case the reciprocal lattice representation is also achiral and instead the chirality, if any, is determined by the distribution of the structure factors (weights) over the reciprocal lattice points. We exemplify these fundamentally different chiralities with two lattices (crystal structures), a lattice of scalene triangles and the snub hexagonal tiling. For each case we discuss how the chirality relates to the reciprocal dual of the crystal and how this reflects on the energy spectrum of the potential designed to self-assemble into the target structure. We also demonstrate the crystallization with snapshots of low energy states self-assembled when simulated in the canonical ensemble.

The simplest chiral  geometric form in two dimensions is the scalene triangle, where the chirality depends on the clockwise order of the side lengths. Hence the simplest chiral crystal is one forming scalene (acute) triangles, an oblique Bravais lattice. 
The oblique lattice is unique in the sense that it is the only chiral Bravais lattice. In this case the chirality of the lattice is directly manifest in the reciprocal lattice and will form scalene triangles with the same chirality as the reciprocal.
Since the basis is trivial there is no freedom in the arrangement of its constituent particles and the only requirement for self-assembly of such a chiral structure is that the positive spectrum $\widehat{V}(k)$ selects for the reciprocal lattice of the target and that the system  has approximately the correct particle density. In Fig.~\ref{paralellogram}(c) a spectrum fulfilling these criteria is shown, together with the corresponding potential and the result of a Monte Carlo simulation with the potential. We see that the mirror symmetry of the system is indeed broken with a homochiral lattice as the result.


As an example of a crystal where the chirality instead emerges from the interplay between the basis and the lattice, we use the snub hexagonal tiling and the crystal formed by its vertices, shown in Fig.~\ref{snubHex}. The tiling together with its dual are the only uniform chiral tilings of the Euclidean plane. This pattern closely resembles geometries observed in experimental systems involving (anisotropic) molecules with a triangular geometry~\cite{Satoshi07}.
An achiral superstructure of vacancies determines the lattice periodicity while the orientation of the basis of six particles in relation to adjacent vacancies determines the chirality. Unlike the previous example, here perturbations of the energy spectrum at $|{\bf G}_i|$ are necessary to distinguish between crystals with different bases, since they consists of more than one particle; see~\cite{edlund_designing_2011} for more details on use of perturbations to break degeneracies of the ground state. 
There are six pairs of reciprocal lattice points representing wave vectors of length $|{\bf G}_i|=\sqrt{7}|{\bf k}_0|$, ${\bf k}_0$ being a primitive reciprocal vector. The Fourier transform of the crystal $\hat{\rho}({\bf k})$ at these sites is equal to $-1$ or $\hat{\rho}_{max}  = 6$. The chirality is determined by the order of the structure factors in the pairwise sites. 
A negative perturbation $-\epsilon$ of the energy spectrum at this $k$ will ensure that the ground state will have the correct basis, as shown in Fig.~\ref{snubHex}(c). 
In the zero temperature limit, a small deformation in the form of a small rotation of the basis by $0.2$ rad towards its chiral counterpart will be visible: the spectrum will not be affected at its maxima by small perturbations of the basis, while a rotation increases the negative energy contribution from the other peaks at that wavelength. 
The maxima correspond to the hexagonal close-packed lattice which, together with the vacancies introduced by the limited density, forms the snub hexagonal tiling.

\begin{figure*}[tb]
\includegraphics[width=0.95 \textwidth]{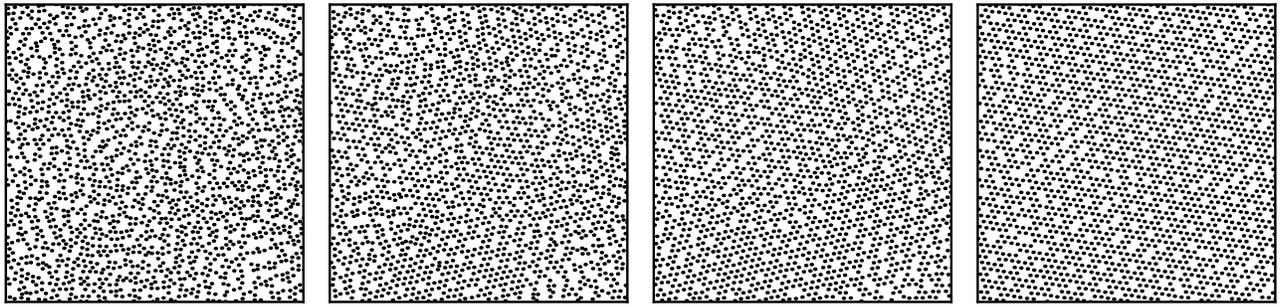}
\caption{ \label{chiralTimeSeries} 
Time evolution of a self-assembling snub hexagonal lattice, after  (from the left) $250$, $2500$, $1.2 \cdot 10^4$ and $10^5$ sweeps of the MC algorithm.}
\end{figure*}

To test whether the target lattice can be assembled from random initial configurations we perform Monte Carlo simulations of particles interacting with the obtained potentials, allowing only local moves and with simulated annealing at fixed density. During the temperature annealing the particles organize into homochiral regions. The boundaries where the chirality changes are energetically disfavored and the homochiral grains grow as the system is annealed towards one of its ground states. The annealing process is however very slow, as illustrated in Fig.~\ref{chiralTimeSeries} where the time evolution is shown on an (approximately) logarithmic time scale. Slow domain growth is typically observed also in experimental systems, which is causing practical difficulties in the synthesis of chiral surfaces~\cite{gellman_chiral_2010}.

\begin{figure}[tb]
\psfrag{a}[c][c][\psLabelSz][0]{\color[rgb]{1,1,1}\bf{a}}
\psfrag{b}[c][c][\psLabelSz][0]{\color[rgb]{1,1,1}\bf{b}}
\psfrag{c}[c][c][\psLabelSz][0]{\color[rgb]{1,1,1}\bf{c}}
\psfrag{d}[c][c][\psLabelSz][0]{\color[rgb]{1,1,1}\bf{d}}
\psfrag{e}[c][c][\psLabelSz][0]{\color[rgb]{1,1,1}\bf{e}}
\psfrag{B}[r][c][\psLabelSz][0]{$\beta$}
\psfrag{L}[c][t][\psLabelSz][0]{$\epsilon$}

\includegraphics[width=0.475 \textwidth]{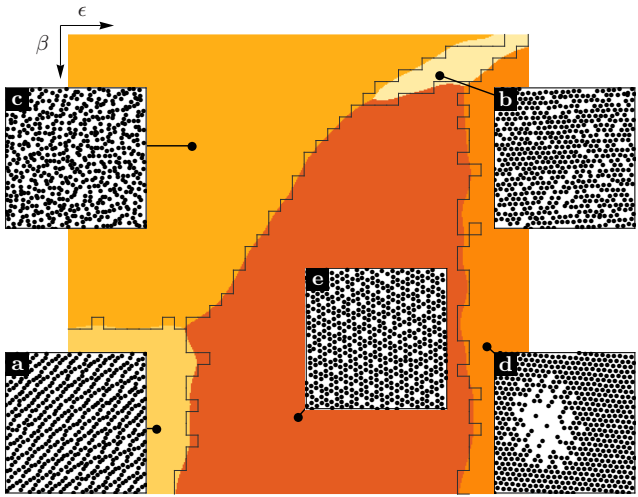}
\caption{ \label{phaseDiagram} 
A phase diagram for the snub hexagonal model shows the equilibrium configuration as a function of inverse temperature $\beta$ and the magnitude $\epsilon$ of the perturbation. Without the perturbation, striped and stripe-like morphologies are observed (a), a strong perturbation leads to vacancy unbinding (b) before complete disorder (c) as temperature increases. A too strong perturbation results in aggregation of the vacancies (d). In the central region (e) the chiral snub hexagonal lattice is formed.}
\end{figure}

Further simulations reveal that the snub hexagonal model has a rich phase diagram, as demonstrated in Fig.~\ref{phaseDiagram}. It shows a large region of stability for the target chiral configuration (c) as temperature and magnitude of the spectrum perturbation $\epsilon$ (see Fig.~\ref{snubHex}) is varied. We can understand the various neighboring phases in terms of the target structure losing some of its properties while retaining others. 
If the negative perturbation stabilizing the basic hexagonal lattice is large, the preference for particles to occupy sites on this lattice will remain even at temperatures where the periodicity of the snub hexagonal lattice, induced by the locations of zeros in the energy spectrum, disappears. The crystal melts through a vacancy unbinding resulting in a hexagonal lattice with randomly placed vacancies, Fig.~\ref{phaseDiagram}(c). For even larger $\epsilon$ the vacancies aggregate into clusters, Fig.~\ref{phaseDiagram}(d), not forming the snub hexagonal lattice for any $\beta$. 
If the negative perturbation is small, the basis will dissolve at temperatures below those where the periodicity of the lattice is broken.  A large variety of basis configurations are observed in this region although the dominant and most stable morphology is striped, Fig.~\ref{phaseDiagram}(b). 
We suspect that the rich phase behavior observed here is typical for models constructed through our design method as the spectrum fixes a set of features of the ground state with varying strengths, each breaking down under different conditions.  This demonstrates that the energy spectrum approach to designing potentials for targeted self-assembly can also be useful when trying to understand and control complex finite temperature phase behavior.

%

In summary, we demonstrate that isotropic pairwise potentials can cause particle systems to self-assemble into chiral surfaces with varying degree of complexity. In general this work hints at some of the future possibilities of nano scale self-assembly, expanding on what we know to be possible, and could be an inspiration for the material science of tomorrow. Lately, there has been remarkable progress on designing nano and colloidal particles with exotic interaction potentials~\cite{likos_effective_2001,min_role_2008}. Still, experimental realizations of systems with as complicated interactions as we use here are not likely to appear in the near future. For theoretical methods and experimental techniques to converge and the field of self-assembly moving from science to industry on a larger scale, further progress from both directions is necessary.

 OL and MNJ acknowledge support from the SuMo Biomaterials center of excellence.  


\bibliographystyle{apsrev4-1}
\bibliography{chiralSurface}

\begin{thebibliography}{23}%
\makeatletter
\providecommand \@ifxundefined [1]{%
 \@ifx{#1\undefined}
}%
\providecommand \@ifnum [1]{%
 \ifnum #1\expandafter \@firstoftwo
 \else \expandafter \@secondoftwo
 \fi
}%
\providecommand \@ifx [1]{%
 \ifx #1\expandafter \@firstoftwo
 \else \expandafter \@secondoftwo
 \fi
}%
\providecommand \natexlab [1]{#1}%
\providecommand \enquote  [1]{``#1''}%
\providecommand \bibnamefont  [1]{#1}%
\providecommand \bibfnamefont [1]{#1}%
\providecommand \citenamefont [1]{#1}%
\providecommand \href@noop [0]{\@secondoftwo}%
\providecommand \href [0]{\begingroup \@sanitize@url \@href}%
\providecommand \@href[1]{\@@startlink{#1}\@@href}%
\providecommand \@@href[1]{\endgroup#1\@@endlink}%
\providecommand \@sanitize@url [0]{\catcode `\\12\catcode `\$12\catcode
  `\&12\catcode `\#12\catcode `\^12\catcode `\_12\catcode `\%12\relax}%
\providecommand \@@startlink[1]{}%
\providecommand \@@endlink[0]{}%
\providecommand \url  [0]{\begingroup\@sanitize@url \@url }%
\providecommand \@url [1]{\endgroup\@href {#1}{\urlprefix }}%
\providecommand \urlprefix  [0]{URL }%
\providecommand \Eprint [0]{\href }%
\providecommand \doibase [0]{http://dx.doi.org/}%
\providecommand \selectlanguage [0]{\@gobble}%
\providecommand \bibinfo  [0]{\@secondoftwo}%
\providecommand \bibfield  [0]{\@secondoftwo}%
\providecommand \translation [1]{[#1]}%
\providecommand \BibitemOpen [0]{}%
\providecommand \bibitemStop [0]{}%
\providecommand \bibitemNoStop [0]{.\EOS\space}%
\providecommand \EOS [0]{\spacefactor3000\relax}%
\providecommand \BibitemShut  [1]{\csname bibitem#1\endcsname}%
\let\auto@bib@innerbib\@empty
\bibitem [{\citenamefont {Okamoto}\ and\ \citenamefont
  {Nakano}(1994)}]{okamoto94}%
  \BibitemOpen
  \bibfield  {author} {\bibinfo {author} {\bibfnamefont {Y.}~\bibnamefont
  {Okamoto}}\ and\ \bibinfo {author} {\bibfnamefont {T.}~\bibnamefont
  {Nakano}},\ }\href@noop {} {\bibfield  {journal} {\bibinfo  {journal} {Chem.
  Rev.}\ }\textbf {\bibinfo {volume} {94}},\ \bibinfo {pages} {349–372}
  (\bibinfo {year} {1994})}\BibitemShut {NoStop}%
\bibitem [{\citenamefont {Eliel}\ and\ \citenamefont {Wilen}(1994)}]{eliel94}%
  \BibitemOpen
  \bibfield  {author} {\bibinfo {author} {\bibfnamefont {E.}~\bibnamefont
  {Eliel}}\ and\ \bibinfo {author} {\bibfnamefont {S.}~\bibnamefont {Wilen}},\
  }\href@noop {} {\emph {\bibinfo {title} {Stereochemistry of Organic
  Compounds}}}\ (\bibinfo  {publisher} {Wileyi-Interscience},\ \bibinfo {year}
  {1994})\BibitemShut {NoStop}%
\bibitem [{\citenamefont {Soai}\ \emph {et~al.}(1995)\citenamefont {Soai},
  \citenamefont {Shibata}, \citenamefont {Morioka},\ and\ \citenamefont
  {Choji}}]{soai95}%
  \BibitemOpen
  \bibfield  {author} {\bibinfo {author} {\bibfnamefont {K.}~\bibnamefont
  {Soai}}, \bibinfo {author} {\bibfnamefont {T.}~\bibnamefont {Shibata}},
  \bibinfo {author} {\bibfnamefont {H.}~\bibnamefont {Morioka}}, \ and\
  \bibinfo {author} {\bibfnamefont {K.}~\bibnamefont {Choji}},\ }\href@noop {}
  {\bibfield  {journal} {\bibinfo  {journal} {Nature}\ }\textbf {\bibinfo
  {volume} {378}},\ \bibinfo {pages} {767} (\bibinfo {year}
  {1995})}\BibitemShut {NoStop}%
\bibitem [{\citenamefont {Soai}\ and\ \citenamefont {Sato}(2002)}]{Soai02}%
  \BibitemOpen
  \bibfield  {author} {\bibinfo {author} {\bibfnamefont {K.}~\bibnamefont
  {Soai}}\ and\ \bibinfo {author} {\bibfnamefont {I.}~\bibnamefont {Sato}},\
  }\href@noop {} {\bibfield  {journal} {\bibinfo  {journal} {Chirality}\
  }\textbf {\bibinfo {volume} {14}},\ \bibinfo {pages} {548} (\bibinfo {year}
  {2002})}\BibitemShut {NoStop}%
\bibitem [{\citenamefont {Edlund}\ \emph
  {et~al.}(2011{\natexlab{a}})\citenamefont {Edlund}, \citenamefont
  {Lindgren},\ and\ \citenamefont {Jacobi}}]{edlund_novel_2011}%
  \BibitemOpen
  \bibfield  {author} {\bibinfo {author} {\bibfnamefont {E.}~\bibnamefont
  {Edlund}}, \bibinfo {author} {\bibfnamefont {O.}~\bibnamefont {Lindgren}}, \
  and\ \bibinfo {author} {\bibfnamefont {M.~N.}\ \bibnamefont {Jacobi}},\
  }\href@noop {} {\bibfield  {journal} {\bibinfo  {journal} {Phys. Rev. Lett.}\
  }\textbf {\bibinfo {volume} {107}},\ \bibinfo {pages} {085501} (\bibinfo
  {year} {2011}{\natexlab{a}})}\BibitemShut {NoStop}%
\bibitem [{\citenamefont {Kondepudi}\ \emph {et~al.}(1990)\citenamefont
  {Kondepudi}, \citenamefont {Kaufman},\ and\ \citenamefont
  {Singh}}]{kondepudi90}%
  \BibitemOpen
  \bibfield  {author} {\bibinfo {author} {\bibfnamefont {D.~K.}\ \bibnamefont
  {Kondepudi}}, \bibinfo {author} {\bibfnamefont {R.}~\bibnamefont {Kaufman}},
  \ and\ \bibinfo {author} {\bibfnamefont {N.}~\bibnamefont {Singh}},\
  }\href@noop {} {\bibfield  {journal} {\bibinfo  {journal} {Science}\ }\textbf
  {\bibinfo {volume} {250}},\ \bibinfo {pages} {975} (\bibinfo {year}
  {1990})}\BibitemShut {NoStop}%
\bibitem [{\citenamefont {Flack}(2003)}]{Flack03}%
  \BibitemOpen
  \bibfield  {author} {\bibinfo {author} {\bibfnamefont {H.}~\bibnamefont
  {Flack}},\ }\href@noop {} {\bibfield  {journal} {\bibinfo  {journal} {Helv.
  Chim. Acta}\ }\textbf {\bibinfo {volume} {86}},\ \bibinfo {pages} {905}
  (\bibinfo {year} {2003})}\BibitemShut {NoStop}%
\bibitem [{\citenamefont {Engel}\ and\ \citenamefont
  {Trebin}(2007)}]{engel_self-assembly_2007}%
  \BibitemOpen
  \bibfield  {author} {\bibinfo {author} {\bibfnamefont {M.}~\bibnamefont
  {Engel}}\ and\ \bibinfo {author} {\bibfnamefont {H.-R.}\ \bibnamefont
  {Trebin}},\ }\href@noop {} {\bibfield  {journal} {\bibinfo  {journal} {Phys.
  Rev. Lett.}\ }\textbf {\bibinfo {volume} {98}},\ \bibinfo {pages} {225505}
  (\bibinfo {year} {2007})}\BibitemShut {NoStop}%
\bibitem [{\citenamefont {Torquato}(2009)}]{torquato_inverse_2009}%
  \BibitemOpen
  \bibfield  {author} {\bibinfo {author} {\bibfnamefont {S.}~\bibnamefont
  {Torquato}},\ }\href@noop {} {\bibfield  {journal} {\bibinfo  {journal} {Soft
  Matter}\ }\textbf {\bibinfo {volume} {5}},\ \bibinfo {pages} {1157} (\bibinfo
  {year} {2009})}\BibitemShut {NoStop}%
\bibitem [{\citenamefont {Blaschke}\ \emph {et~al.}(1979)\citenamefont
  {Blaschke}, \citenamefont {Kraft}, \citenamefont {Fickentscher},\ and\
  \citenamefont {K\"ohler}}]{blaschke_chromatographic_1979}%
  \BibitemOpen
  \bibfield  {author} {\bibinfo {author} {\bibfnamefont {G.}~\bibnamefont
  {Blaschke}}, \bibinfo {author} {\bibfnamefont {H.}~\bibnamefont {Kraft}},
  \bibinfo {author} {\bibfnamefont {K.}~\bibnamefont {Fickentscher}}, \ and\
  \bibinfo {author} {\bibfnamefont {F.}~\bibnamefont {K\"ohler}},\ }\href@noop
  {} {\bibfield  {journal} {\bibinfo  {journal} {{Arznei.-Forschung}}\ }\textbf
  {\bibinfo {volume} {29}},\ \bibinfo {pages} {1640} (\bibinfo {year}
  {1979})}\BibitemShut {NoStop}%
\bibitem [{\citenamefont {Yoon}\ and\ \citenamefont
  {Jacobsen}(2003)}]{yoon_privileged_2003}%
  \BibitemOpen
  \bibfield  {author} {\bibinfo {author} {\bibfnamefont {T.~P.}\ \bibnamefont
  {Yoon}}\ and\ \bibinfo {author} {\bibfnamefont {E.~N.}\ \bibnamefont
  {Jacobsen}},\ }\href@noop {} {\bibfield  {journal} {\bibinfo  {journal}
  {Science}\ }\textbf {\bibinfo {volume} {299}},\ \bibinfo {pages} {1691 }
  (\bibinfo {year} {2003})}\BibitemShut {NoStop}%
\bibitem [{\citenamefont {Jacobsen}\ \emph {et~al.}(1999)\citenamefont
  {Jacobsen}, \citenamefont {Pfaltz},\ and\ \citenamefont
  {Yamamoto}}]{jacobsen_comprehensive_1999}%
  \BibitemOpen
  \bibfield  {author} {\bibinfo {author} {\bibfnamefont {E.~N.}\ \bibnamefont
  {Jacobsen}}, \bibinfo {author} {\bibfnamefont {A.}~\bibnamefont {Pfaltz}}, \
  and\ \bibinfo {author} {\bibfnamefont {H.}~\bibnamefont {Yamamoto}},\
  }\href@noop {} {\emph {\bibinfo {title} {Comprehensive Asymmetric
  Catalysis}}},\ Vol.\ \bibinfo {volume} {1--3}\ (\bibinfo  {publisher}
  {Springer, Berlin},\ \bibinfo {year} {1999})\BibitemShut {NoStop}%
\bibitem [{\citenamefont {Schunack}\ \emph {et~al.}(2001)\citenamefont
  {Schunack}, \citenamefont {Petersen}, \citenamefont {K\"{u}hnle},
  \citenamefont {L{\ae}gsgaard}, \citenamefont {Stensgaard}, \citenamefont
  {Johannsen},\ and\ \citenamefont {Besenbacher}}]{schunack_anchoring_2001}%
  \BibitemOpen
  \bibfield  {author} {\bibinfo {author} {\bibfnamefont {M.}~\bibnamefont
  {Schunack}}, \bibinfo {author} {\bibfnamefont {L.}~\bibnamefont {Petersen}},
  \bibinfo {author} {\bibfnamefont {A.}~\bibnamefont {K\"{u}hnle}}, \bibinfo
  {author} {\bibfnamefont {E.}~\bibnamefont {L{\ae}gsgaard}}, \bibinfo {author}
  {\bibfnamefont {I.}~\bibnamefont {Stensgaard}}, \bibinfo {author}
  {\bibfnamefont {I.}~\bibnamefont {Johannsen}}, \ and\ \bibinfo {author}
  {\bibfnamefont {F.}~\bibnamefont {Besenbacher}},\ }\href@noop {} {\bibfield
  {journal} {\bibinfo  {journal} {Phys. Rev. Lett.}\ }\textbf {\bibinfo
  {volume} {86}},\ \bibinfo {pages} {456} (\bibinfo {year} {2001})}\BibitemShut
  {NoStop}%
\bibitem [{\citenamefont {Sholl}\ \emph {et~al.}(2001)\citenamefont {Sholl},
  \citenamefont {Asthagiri},\ and\ \citenamefont
  {Power}}]{sholl_naturally_2001}%
  \BibitemOpen
  \bibfield  {author} {\bibinfo {author} {\bibfnamefont {D.~S.}\ \bibnamefont
  {Sholl}}, \bibinfo {author} {\bibfnamefont {A.}~\bibnamefont {Asthagiri}}, \
  and\ \bibinfo {author} {\bibfnamefont {T.~D.}\ \bibnamefont {Power}},\
  }\href@noop {} {\bibfield  {journal} {\bibinfo  {journal} {J. Phys. Chem. B}\
  }\textbf {\bibinfo {volume} {105}},\ \bibinfo {pages} {4771} (\bibinfo {year}
  {2001})}\BibitemShut {NoStop}%
\bibitem [{\citenamefont {Lorenzo}\ \emph {et~al.}(2000)\citenamefont
  {Lorenzo}, \citenamefont {Baddeley}, \citenamefont {Muryn},\ and\
  \citenamefont {Raval}}]{ortega_lorenzo_extended_2000}%
  \BibitemOpen
  \bibfield  {author} {\bibinfo {author} {\bibfnamefont {M.~O.}\ \bibnamefont
  {Lorenzo}}, \bibinfo {author} {\bibfnamefont {C.~J.}\ \bibnamefont
  {Baddeley}}, \bibinfo {author} {\bibfnamefont {C.}~\bibnamefont {Muryn}}, \
  and\ \bibinfo {author} {\bibfnamefont {R.}~\bibnamefont {Raval}},\
  }\href@noop {} {\bibfield  {journal} {\bibinfo  {journal} {Nature}\ }\textbf
  {\bibinfo {volume} {404}},\ \bibinfo {pages} {376} (\bibinfo {year}
  {2000})}\BibitemShut {NoStop}%
\bibitem [{\citenamefont {Spillmann}\ \emph {et~al.}(2003)\citenamefont
  {Spillmann}, \citenamefont {Dmitriev}, \citenamefont {Lin}, \citenamefont
  {Messina}, \citenamefont {Barth},\ and\ \citenamefont
  {Kern}}]{spillmann_hierarchical_2003}%
  \BibitemOpen
  \bibfield  {author} {\bibinfo {author} {\bibfnamefont {H.}~\bibnamefont
  {Spillmann}}, \bibinfo {author} {\bibfnamefont {A.}~\bibnamefont {Dmitriev}},
  \bibinfo {author} {\bibfnamefont {N.}~\bibnamefont {Lin}}, \bibinfo {author}
  {\bibfnamefont {P.}~\bibnamefont {Messina}}, \bibinfo {author} {\bibfnamefont
  {J.~V.}\ \bibnamefont {Barth}}, \ and\ \bibinfo {author} {\bibfnamefont
  {K.}~\bibnamefont {Kern}},\ }\href@noop {} {\bibfield  {journal} {\bibinfo
  {journal} {J. Am. Chem. Soc.}\ }\textbf {\bibinfo {volume} {125}},\ \bibinfo
  {pages} {10725} (\bibinfo {year} {2003})}\BibitemShut {NoStop}%
\bibitem [{\citenamefont {Gellman}(2010)}]{gellman_chiral_2010}%
  \BibitemOpen
  \bibfield  {author} {\bibinfo {author} {\bibfnamefont {A.~J.}\ \bibnamefont
  {Gellman}},\ }\href@noop {} {\bibfield  {journal} {\bibinfo  {journal} {ACS
  Nano}\ }\textbf {\bibinfo {volume} {4}},\ \bibinfo {pages} {5} (\bibinfo
  {year} {2010})}\BibitemShut {NoStop}%
\bibitem [{\citenamefont {Edlund}\ and\ \citenamefont
  {Jacobi}(2010)}]{edlund_universality_2010}%
  \BibitemOpen
  \bibfield  {author} {\bibinfo {author} {\bibfnamefont {E.}~\bibnamefont
  {Edlund}}\ and\ \bibinfo {author} {\bibfnamefont {M.~N.}\ \bibnamefont
  {Jacobi}},\ }\href@noop {} {\bibfield  {journal} {\bibinfo  {journal} {Phys.
  Rev. Lett.}\ }\textbf {\bibinfo {volume} {105}},\ \bibinfo {pages} {137203}
  (\bibinfo {year} {2010})}\BibitemShut {NoStop}%
\bibitem [{\citenamefont {Folland}(1992)}]{folland_fourier_2009}%
  \BibitemOpen
  \bibfield  {author} {\bibinfo {author} {\bibfnamefont {G.~B.}\ \bibnamefont
  {Folland}},\ }\enquote {\bibinfo {title} {{Fourier Analysis and Its
  Applications}},}\ \ (\bibinfo  {publisher} {Brooks/Cole Pub Co},\ \bibinfo
  {address} {Pacific Grove},\ \bibinfo {year} {1992})\ Chap.\ \bibinfo
  {chapter} {7.5}, p.\ \bibinfo {pages} {246}\BibitemShut {NoStop}%
\bibitem [{\citenamefont {Edlund}\ \emph
  {et~al.}(2011{\natexlab{b}})\citenamefont {Edlund}, \citenamefont
  {Lindgren},\ and\ \citenamefont {Jacobi}}]{edlund_designing_2011}%
  \BibitemOpen
  \bibfield  {author} {\bibinfo {author} {\bibfnamefont {E.}~\bibnamefont
  {Edlund}}, \bibinfo {author} {\bibfnamefont {O.}~\bibnamefont {Lindgren}}, \
  and\ \bibinfo {author} {\bibfnamefont {M.~N.}\ \bibnamefont {Jacobi}},\
  }\href@noop {} {\bibfield  {journal} {\bibinfo  {journal} {Phys. Rev. Lett.}\
  }\textbf {\bibinfo {volume} {107}},\ \bibinfo {pages} {085503} (\bibinfo
  {year} {2011}{\natexlab{b}})}\BibitemShut {NoStop}%
\bibitem [{\citenamefont {Katano}\ \emph {et~al.}(2007)\citenamefont {Katano},
  \citenamefont {Kim}, \citenamefont {Matsubara}, \citenamefont {Kitagawa},\
  and\ \citenamefont {Kawai}}]{Satoshi07}%
  \BibitemOpen
  \bibfield  {author} {\bibinfo {author} {\bibfnamefont {S.}~\bibnamefont
  {Katano}}, \bibinfo {author} {\bibfnamefont {Y.}~\bibnamefont {Kim}},
  \bibinfo {author} {\bibfnamefont {H.}~\bibnamefont {Matsubara}}, \bibinfo
  {author} {\bibfnamefont {T.}~\bibnamefont {Kitagawa}}, \ and\ \bibinfo
  {author} {\bibfnamefont {M.}~\bibnamefont {Kawai}},\ }\href@noop {}
  {\bibfield  {journal} {\bibinfo  {journal} {J. Am. Chem. Soc.}\ }\textbf
  {\bibinfo {volume} {129}},\ \bibinfo {pages} {2511} (\bibinfo {year}
  {2007})}\BibitemShut {NoStop}%
\bibitem [{\citenamefont {Likos}(2001)}]{likos_effective_2001}%
  \BibitemOpen
  \bibfield  {author} {\bibinfo {author} {\bibfnamefont {C.~N.}\ \bibnamefont
  {Likos}},\ }\href@noop {} {\bibfield  {journal} {\bibinfo  {journal} {Phys.
  Rep.}\ }\textbf {\bibinfo {volume} {348}},\ \bibinfo {pages} {267} (\bibinfo
  {year} {2001})}\BibitemShut {NoStop}%
\bibitem [{\citenamefont {Min}\ \emph {et~al.}(2008)\citenamefont {Min},
  \citenamefont {Akbulut}, \citenamefont {Kristiansen}, \citenamefont {Golan},\
  and\ \citenamefont {Israelachvili}}]{min_role_2008}%
  \BibitemOpen
  \bibfield  {author} {\bibinfo {author} {\bibfnamefont {Y.}~\bibnamefont
  {Min}}, \bibinfo {author} {\bibfnamefont {M.}~\bibnamefont {Akbulut}},
  \bibinfo {author} {\bibfnamefont {K.}~\bibnamefont {Kristiansen}}, \bibinfo
  {author} {\bibfnamefont {Y.}~\bibnamefont {Golan}}, \ and\ \bibinfo {author}
  {\bibfnamefont {J.}~\bibnamefont {Israelachvili}},\ }\href@noop {} {\bibfield
   {journal} {\bibinfo  {journal} {Nat. Mater.}\ }\textbf {\bibinfo {volume}
  {7}},\ \bibinfo {pages} {527} (\bibinfo {year} {2008})}\BibitemShut {NoStop}%
\end{thebibliography}%

\end{document}